\documentstyle{mn}

\title{A Search for $\gamma$ Doradus-Type Variable Stars in the Hyades}
\author[K. Krisciunas et al.]
       {Kevin Krisciunas,$^1$ Richard A. Crowe,$^2$
        Kenneth D. Luedeke$^3$ and Mavourneen Roberts$^2$  \\
      $^1$Joint Astronomy Centre, 660 N. A`oh\={o}k\={u} Place, University
Park, Hilo, HI 96720, USA \\
      $^2$Department of Physics and Astronomy, University of Hawaii
at Hilo, 200 West Kawili Street, Hilo, Hawaii 96720, USA \\
      $^3$9624 Giddings Avenue NE, Albuquerque, NM 87109, USA}
\date{Accepted 1995 July 12}

\begin{document}

\maketitle

\begin {abstract}
$\gamma$ Doradus stars are photometrically variable early F-type stars
on or just above the main sequence in the Hertzsprung-Russell Diagram.
They vary by as much as 0.1 mag on time scales an order of magnitude
slower than the fundamental radial pulsation period.  These brightness
variations are presumably due to non-radial gravity-mode pulsations.
We obtained high precision $V$-band photometry of 8 F0 to F5 stars in the
Hyades
(BS 1319, BS 1354, BS 1385, BS 1408, BS 1430, BS 1432, BS 1459, and BS 1472)
and found that none of them showed strong evidence of
$\gamma$ Dor-type variability.  Since $\gamma$ Dor-type candidates are
found in the Pleiades and in NGC 2516 (having ages of 78 and 137 Myr,
repectively) but apparently not in the Hyades (age $>$ 600 Myr), we
hypothesize that the $\gamma$ Dor phenomenon is a characteristic of
relatively young F stars.

Two of the stars investigated showed marginal evidence of low-amplitude
variability.  The $\pm$3 mmag variability of the F5 star BS 1319
is most likely due to rotational modulation of star spots, though it is
not impossible that it is a $\gamma$ Dor star.  Another F5 star BS 1459
($\Delta V = \pm 2$ mmag) has a possible period similar to $\delta$ Scuti
stars, but no firm conclusions should be made concerning its behavior unless
and until its variability is confirmed.

\end{abstract}

\section{Introduction}

$\delta$ Scuti stars, SX Phoenicis stars, and $\gamma$ Doradus stars are three
kinds of pulsating variable stars of similar spectral type and luminosity
class.  All three types are found near the main sequence in (or near) the
Cepheid instability strip in the Hertzsprung-Russell (HR) Diagram.

$\delta$ Scuti stars comprise the largest class of these three types of stars
(Breger 1979, Rodriguez et al. 1994), and are subsequently the best studied.
They typically have
photometric periods of 1--3 h, low photometric amplitudes (0.02 mag), and
have metal abundances and space motions typical of Population I stars.

SX Phoenicis stars have similar periods but
typically have much higher photometric amplitudes
(0.3 to 0.8 mag).  They have metal abundances and space motions typical
of Population II stars, and are blue stragglers in the post-main sequence
stage of evolution.  (See Garcia et al. 1990; Kim, McNamara \& Christensen
1993.)

$\gamma$ Doradus stars are the most recently identified of the three
types of variable stars just mentioned.  They are typically
early-F stars on, or just above, the main sequence in the HR Diagram,
and they are at or beyond the cool edge of the Cepheid
instability strip.  They exhibit photometric variability as large as
0.1 mag in $V$ on a time scale of 0.5 to 3 d.
At the time of this writing there were 17 known candidates of this type
(Krisciunas \& Handler 1995).
The best studied examples are $\gamma$ Doradus itself (Balona, Krisciunas \&
Cousins 1994), 9 Aurigae (Krisciunas et al. 1995a), HD 224638 and HD 224945
(Mantegazza, Poretti \& Zerbi 1994).

$\delta$ Scuti stars and SX Phoenicis stars are believed to exhibit
radial pulsations (in the fundamental mode),
radial overtones, and non-radial pressure mode ($p$-mode)
oscillations.  Recently, Breger (1993) has shown that some $\delta$
Scuti stars also show evidence of lower frequency (longer period) variations,
which one would attribute to non-radial gravity mode ($g$-mode) pulsations.
Amongst $\gamma$ Doradus stars, $\gamma$ Dor and 9 Aur
have photometric periods, radial velocity variations,
and line profile variations consistent with the notion that they are
exhibiting $g$-mode pulsations.  Recently, Aerts \& Krisciunas (1995) have
found that 9 Aur can be modelled as an $\ell=3, |m|=1$ non-radial pulsator.

An interesting question one might pose is: how is the pulsational behavior
outlined above related to the age of A- and F-type stars?  We know that
SX Phe stars are ``old'', typically 1--2 Gyr for a mass of 1.6 M$_{\odot}$
(Rodriguez, Rolland \& Lopez de Coca 1990).  $\gamma$ Dor
itself is imbedded in a $\beta$ Pictoris-type disk or envelope (Walker \&
Wolstencroft 1988), and so is presumably ``young''.
Antonello and Mantegazza (1986) found four
``slowly'' varying F stars in the cluster NGC 2516.  Mantegazza (private
communication) indicates that subsequent unpublished photometry of this
cluster revealed other candidates.  The main sequence turnoff point for
NGC 2516 corresponds to a spectral type of a late-B star, so its age
should be comparable to the Pleiades (i.e. 78 Myr; Lang 1992).  We note that
the Pleiades itself has one $\gamma$ Dor candidate, HD 23375 (Breger 1972).

It would be interesting to determine if $\gamma$ Dor stars, $\delta$ Scuti
stars and SX Phe stars are to be found only in certain ranges of age.
Since we know of only a relatively small number of $\gamma$ Dor stars, it would
also be useful to increase the number of known candidates.
In this paper we report our attempts to identify $\gamma$ Dor stars in the
Hyades, the age of which is 661 Myr (Lang 1992).

\section{Observations}

We chose stars in the Hyades from Welch's (1979) list and considered
information given in the The Bright Star Catalogue (Hoffleit \& Jaschek 1982).
Known $\delta$ Scuti stars were avoided.  This left us with 8 early- to mid-F
stars.  To our knowledge none of these 8 stars has ever been checked
carefully for variability on the time scale we are interested in here.
However, we note that three of our stars (BS 1430, 1432, and 1472) were
observed by Horan (1979) in his search for $\delta$ Scuti stars in the
Hyades.  From 3 to 4 hours of monitoring, he found that each of these three
stars was constant to 3 mmag or better.  In order to find out if any of
these stars showed $\gamma$ Dor-like variability, observations over longer
time scales were still warranted.

Our observations were principally made with the 0.6-m reflector at Mauna
Kea, Hawaii, using an Optec SSP-5 photometer owned by the University of
Hawaii at Hilo.  Other measurements were made by one of us (KK) at Cerro
Tololo,
Chile, using the 0.6-m Lowell telescope, a dry ice cooled photomultiplier
tube in the ASCAP photometer, and a 4.5 mag neutral density filter.  Luedeke's
data were obtained with his 0.25-m reflector in Albuquerque, New Mexico,
and an Optec SSP-3 photometer.  Appropriate $V$-band differential
extinctions corrections were made with values of the extinction derived
nightly.  Transformation to the $UBV$ photometric system was accomplished
with observations of red-blue pairs of known brightness and color
(Hall 1983).

The Mauna Kea and Cerro Tololo data were taken in the following manner:
comparison star -- two program stars -- comparison star -- two other
program stars -- comparison star, etc.
One cycle on the 8 program stars, bracketed by
5 measures of the comparison star, took between 35 and 50 minutes.  (This was
mostly owing to the constraint of having to set the telescope manually.)
Each measure of a star actually consisted of two consecutive 30 second
integrations.  If these two measures were not within one percent, we often
found that the star was no longer centered in the diaphragm, or the seeing
had gone bad, or the dome slit was in the way.  Then a third or a fourth
30 second integration was taken.

We used BS 1422 (=vB 80) as our comparison star.  It is a spectroscopic
binary comprised of an F0V and a G2V star.
This comparison star was suggested by Handler (private
communication), who used it for over 100 h of photometry in another
photometric project.  Given that many of the Hyades program stars proved to
be constant within the errors of observation (see below), the choice of
BS 1422 as the comparison star turned out to be an excellent one.

In Table 1 we give a summary of the {\it V}-band photometry, which was
obtained from 1994 November 15 to 1995 January 23 UT.
Typically, an individual differential magnitude obtained with
the 0.6-m telescopes had an internal error of $\pm$ 4 mmag.
For Luedeke's data a typical individual measure is good to $\pm$ 10 mmag.
To equalize the weighting for subsequent analysis, each of Luedeke's data
points represents the mean of three differential measures.
The data can be obtained from IAU Commission 27 as file 302E of unpublished
photometry.  (See Breger, Jaschek \& Dubois 1990 for further information.)

\begin{table}
\caption{Summary of differential photometry of Hyades stars.  The comparison
star in all cases was BS 1422 (= vB 80; $V = 5.58$, $B$--$V = 0.32$).
For each star we give the catalogue number
from The Bright Star Catalogue, the corresponding number in van
Bueren's (1952) list, the spectral type, the mean differential
$V$ magnitude, the internal error of a single differential value
(i.e. the standard deviation of the distribution, {\em not} the mean error
of the mean) the number of nights on which data were taken, and the number
of data points.}
\begin{tabular}{rrlrcrr}\hline
BS & vB & Sp & $<\Delta V>$ & $\sigma$(mmag) & N  & n \\
\hline
  $1319$  &  $ 20$ & $\rm{F5V }$ & $ 0.7410$ & $3.7$ & $20$ & $ 76$  \\
  $1354$  &  $ 32$ & $\rm{F3:V}$ & $ 0.5372$ & $5.1$ & $18$ & $ 45$  \\
  $1385$  &  $ 53$ & $\rm{F4V }$ & $ 0.4094$ & $4.5$ & $18$ & $ 44$  \\
  $1408$  &  $ 68$ & $\rm{F0IV}$ & $ 0.3362$ & $3.2$ & $19$ & $ 60$  \\
  $1430$  &  $ 84$ & $\rm{F0V }$ & $-0.1639$ & $3.5$ & $18$ & $ 45$  \\
  $1432$  &  $ 89$ & $\rm{F4V }$ & $ 0.4397$ & $4.0$ & $18$ & $ 45$  \\
  $1459$  &  $100$ & $\rm{F5IV}$ & $ 0.4482$ & $3.7$ & $20$ & $102$  \\
  $1472$  &  $103$ & $\rm{F0V }$ & $ 0.2180$ & $3.5$ & $19$ & $ 61$  \\
\hline
\end{tabular}
\end{table}

The reader will notice that the average number of data points obtained per
night (per star) was small.  If any of the stars listed in Table 1
were photometrically variable in the manner of $\gamma$ Dor or 9 Aur, they
would easily have flagged themselves as ``stars worthy of further
investigation''.  We assert this on the basis of actual photometry of
$\gamma$ Dor and 9 Aur obtained during the 10-night observing run at Cerro
Tololo and the 7-night observing run at Mauna Kea.  We took subsets of
photometry of $\gamma$ Dor and 9 Aur, amounting to 4 points per night.
The range of the data in both cases was nearly 0.08 mag (giving evidence
of variability at the 15-$\sigma$ level or better).  Power spectra of these
subsets of data revealed the principal frequency plus some aliases.  False
alarm probabilities of the highest peaks in the power spectra of such
subsets of data ranged from $3 \times 10^{-4}$ to $4 \times 10^{-3}$.

It was our expectation that some of the Hyades stars would turn out to be
very constant.\footnote[1] {It could be argued that a star is either
``constant'' or ``variable'' and cannot be ``very constant'' or
``somewhat constant''.  However, it is easier to determine that a
variable star is
variable than to prove that a constant star is constant, because in the latter
case one must cover all reasonable time scales, from minutes to years,
and one always has to deal with photon statistics.  Here
we define a {\em constant} star to be: (1) one that shows very little range in
the data values (i.e. $\pm$2-$\sigma$, where $\sigma$ is the typical
internal error of a single differential measurement; and (2) one for which
the power spectrum of the
photometry shows no significant peaks.  The definition
of ``significant peak'' must be left somewhat vague, because it depends on
the mechanism of variability of a star and the coherency argument relating
to the reality of a given frequency.  For example, we recently found
that the star 7 Cam (= BS 1568) is a low-amplitude ($\pm$6 mmag),
photometrically variable, ellipsoidal star.  (It was observed during the
7 night run at Mauna Kea, during which most of the data of this paper were
obtained.)  The only significant peak in the power spectrum of its photometry
corresponds to half the orbital period of the star about its unseen companion.
(See Krisciunas et al. 1995b.) In this paper we will argue for the significance
of one peak in the power spectrum of the photometry of BS 1319, owing to
its projected rotational velocity.}  That would give us an estimate of
the accuracy of the photometry.  If any of the 8 Hyades stars were
found to be constant with respect to BS 1422, then we would know that BS 1422
is a ``safe'' comparison star.  Suspected variability of another star,
observed differentially with respect to BS 1422, would be safely attributed to
the program star.

\section{Analysis of photometry}

 \begin{figure*}
 \vspace*{90mm}
 \caption{Power spectrum of the $V$-band photometry of BS 1385 vs.
BS 1422, obtained at Mauna Kea from 1995 December 29 to 1995 January 4 UT.}
 \end{figure*}

Our principal tool for the analysis of the photometry was to produce
power spectra, using the Lomb-Scargle algorithm as presented by Press
\& Teukolsky (1988).  Typically, we chose an oversampling factor of 4
and calculated the power spectra to a frequency two times the Nyquist
frequency ({\sc hifac} = 2).  It is allowable to investigate frequencies beyond
the Nyquist frequency if one has unequally sampled data, as is always the
case with astronomical photometry carried out over a number of days
primarily at a single site.  Indeed, in the case of the low-amplitude
ellipsoidal star BS 1568, observations by Guinan and McCook from 1989/90 reveal
the true frequency of variability only if one calculates the power spectrum
to 2.3 times the Nyquist frequency (Krisciunas et al. 1995b).

One of the parameters obtained from the analysis is the
false alarm probability of any given peak in the power spectrum.
This is the probability that a {\em random} sample would give a peak of a given
height.  Large data sets on periodically variable stars such as $\gamma$ Dor
give vanishingly small false alarm probabilities (e.g. 10$^{-20}$).
False alarm probabilities of 0.01 are usually significant.

To refine the value of the frequency of a peak indicated by the power
spectrum, we used the multiple period determination program {\sc perdet}
(Breger 1989).  This also allowed a least-squares determination of the
amplitude and phase of the sinusoid.

During our 10 night observing run (1994 November) at Cerro Tololo as part
of a multi-longitude campaign on $\gamma$ Dor, we observed each of stars
in Table 1 a total of 17 times when the Hyades was near the meridian.
Luedeke also observed all 8 stars on one
night just prior, and one night just after, the CTIO observations.  Clearly,
the Hyades cluster is not well placed for observations at 30 degrees south
latitude, but on the basis of this small amount of data we already knew that
the Hyades F-type stars were constant to $\pm$0.01 mag or better.

Mauna Kea is at an ideal latitude for Hyades observations.  Over the course
of a 7 night multi-longitude campaign on 9 Aur (the northern prototypical
$\gamma$ Dor star), we observed the Hyades stars again and
obtained typically 4 points per night.  (Except for New Year's Eve, when
we were hampered by winds, the observing conditions were excellent.)
As mentioned above, if any of these  stars were exhibiting $\gamma$ Dor-type
activity, that would have been revealed.

Fig 1 shows an altogether uninteresting power spectrum of the
BS 1385 vs. BS 1422 data obtained at Mauna Kea.
The largest peak in the power spectrum gives a false alarm probability
of 0.78.  There is clearly no evidence of periodic variability.

 \begin{figure*}
 \vspace*{90mm}
 \caption{Power spectrum of all available photometry of BS 1319 vs. BS 1422
(1994 November 15 to 1995 January 23 UT).}
 \end{figure*}

 \begin{figure*}
 \vspace*{90mm}
 \caption{All available photometry of BS 1319 vs. BS 1422, folded with
a period of 1.4336 d.}
 \end{figure*}

{}From the internal errors listed in column 5 of Table 1 and power spectra
such as that shown in Fig 1 we conclude that the following stars exhibit no
evidence of variability: BS 1354, BS 1385, BS 1408, BS 1430, BS 1432, and
BS 1472.

Power spectra of the photometry of BS 1319 and BS 1459 gave hints that
they might be low-amplitude variables.
In the case of BS 1319 the suggested period was 1.4 d.  In the
case of BS 1459 the suggested period was in the range 3--5 h.

On the nights of 1995 January 22 and 23 UT we concentrated our efforts
on these two stars, obtaining 57 points on BS 1319 and 31 points on BS
1459.  The comparison star once again was BS 1422.  BS 1408 was used as
the check star on January 22.  BS 1472 was used as the check star on
January 23.

Fig 2 shows the power spectrum of all the available data on BS 1319
(1994 November 15 to 1995 January 23 UT).
We believe that $f$ = 0.6976 d$^{-1}$ is a true frequency.  (The
other peak would simply be its one-day alias).  The false
alarm probability of this peak is 0.02.

Fig 3 shows all the available BS 1319 data folded with a period of
1/$f$ = 1.4336 d.  The sinusoid shown has an amplitude of 3.0 mmag.
That amplitude has a 1-$\sigma$ error of $\pm$0.5 mmag.

 \begin{figure*}
 \vspace*{90mm}
 \caption{Power spectrum of BS 1459 vs. BS 1422 photometry obtained at
Mauna Kea on 1995 January 22 and 23 UT.}
 \end{figure*}

 \begin{figure*}
 \vspace*{90mm}
 \caption{BS 1459 vs. BS 1422 data obtained on 1995 January 22 and 23 UT.
The data have been folded with a period of 0.1670 d.}
 \end{figure*}

Fig 4 shows the power spectrum of the BS 1459 vs. BS 1422 data obtained
on the nights of 1995 January 22 and 23 UT.
The false alarm probability of the highest peak in the power spectrum is
0.28.

Fig 5 shows the photometry of BS 1459 vs. BS 1422 obtained on 1995 January
22 and 23 UT, folded with a period of 0.1670 d.  The sinusoid shown has an
amplitude of 1.8 mmag.  That amplitude has a 1-$\sigma$ error of $\pm$0.5 mmag.

\section{Two low-amplitude variables?}

We believe BS 1319 to be variable at the $\pm$3 mmag
level, and we believe this to be due to rotational modulation.  We have
calculated the radius of the star using
Str$\rm \ddot{o}$mgren photometry (Hauck and Mermilliod 1990)
as input.  The star's absolute magnitude and radius are
calculated by means of relationships given by Balona \& Shobbrook (1984)
and Moon (1984).  The resulting size of the star is 1.53 R$_{\odot}$.
Given its projected rotational velocity of $v$ sin $i$ = 53 km sec$^{-1}$
(Hoffleit \& Jaschek 1982),
the derived rotational period is 1.46 d times sin $i$.  If we are looking
at the star side-on ($i \approx 90^{\circ}$), this closely approximates the
period of 1.4336 d obtained from the photometry.

Because of the near coincidence of the corresponding period of one of the
two peaks in the power spectrum given in Fig 2 with the maximum allowed
rotational period of the star,
the simplest explanation for the variability of BS 1319 would be rotational
modulation, presumably of star spots.  Now, it is generally believed that
stars with spectral types earlier than
F7 do not show evidence of star spots (Giampapa and Rosner 1984). However,
G$\rm \ddot{u}$del, Schmitt \& Benz (1995) report the surprising result
that the F0V star 47 Cas exhibits evidence for strong coronal activity.
Spotted stars often have enhanced coronal activity.
The kind of variability hinted at in our photometry of BS 1319 is two
orders of magnitude less than that found on some spotted stars.

While the coherency argument favors the star spot hypothesis for
the variability, it is {\em possible} that the BS 1319 is a
$\gamma$ Dor-type star.  To prove this would require extensive and highly
accurate photometry carried out during a multi-site campaign.  One
would be looking for evidence of multiple periods,
as are found in other $\gamma$ Dor candidates.  Given the small photometric
range of BS 1319 ($\Delta V < 0.01$ mag vs. $\approx$ 0.1 mag for 9 Aur and
$\gamma$ Dor), the other signatures of pulsation such as radial velocity
and line profiles changes would probably not be measurable.

Regarding BS 1459, it is {\em possible} that it is variable at the $\pm$2 mmag
level, but demonstrating this convincingly is another matter.  Given the
relatively high false alarm probability of the highest peak in the power
spectrum of Fig 4, and the low amplitude of the best sinusoid shown in Fig 5,
we are not confident in saying that BS 1459 is definitely variable.

\section{Comparison with known $\delta$ Scuti stars in the Hyades}

Let us assume for the sake or argument that both BS 1319 and BS 1459 are
pulsating.  How do their pulsation constants compare to known $\delta$
Scuti stars in the Hyades?

Rodriguez et al. (1994) give an extensive catalogue of $\delta$ Scuti stars,
8 of which (BS 1351, 1356, 1368, 1392, 1394, 1412, 1444, and 1547) are in
the Hyades.  We have calculated the pulsation constants $Q$ from the equation
given by Breger \& Bregman (1975):

\begin{equation}
{\rm log}\ Q = -6.454 + {\rm log}\ P + 0.5\ {\rm log}\ g + 0.1\ M_{bol} +
{\rm log}\ T_{eff} .
\end{equation}

Values of log\ $g$ and $T_{eff}$ were calculated from the Str$\rm
\ddot{o}$mgren
photometry given by Rodriguez et al. (1994) and software based on Moon
\& Dworetsky (1985).  The absolute
visual magnitudes were also calculated from the Str$\rm \ddot{o}$mgren
photometry, rather than from the apparent magnitudes, slight reddening
corrections, and the Hyades distance modulus.  We adopted a bolometric
correction for stars of this spectral type of -0.10 (Harris 1963).

In Fig 6 we plot a color-magnitude diagram containing 16 of the 17
$\gamma$ Dor candidates given by Krisciunas \& Handler (1995), the 8
$\delta$ Scuti stars in the Hyades, and the two low-amplitude variables from
this paper.  We note that the bona fide $\gamma$ Dor stars (plotted as circles
in Fig 6) are found in a very small region of the diagram, near the
intersection of the cool edge of the instability strip and the main sequence,
while the $\delta$ Scuti stars are distributed throughout the instability
strip.

 \begin{figure*}
 \vspace*{90mm}
 \caption{A color-magnitude diagram for stars discussed in this paper.
Circles: bona fide $\gamma$ Dor stars from Krisciunas \& Handler (1995).
Triangles: $\gamma$ Dor candidates in NGC 2516.  Dots: other $\gamma$ Dor
candidates given by Krisciunas \& Handler (1995). Squares: $\delta$
Scuti stars listed by Rodriguez et al. (1994). Asterisk: BS 1459.  Five
pointed star: BS 1319.  The zero age main sequence is adopted from
Crawford (1975, 1979), while the borders of the Cepheid instability strip
are taken from Breger (1979).}
 \end{figure*}

For the 8 $\delta$ Scuti stars we find $0.016 < Q < 0.031$ d.  The larger
values of $Q$ are consistent with these stars exhibiting pulsation in the
fundamental radial mode or in the first radial overtone, while the smaller
values of $Q$ might correspond to higher radial overtones (Stellingwerf 1979).
However, we would like to emphasize that {\em one cannot determine the
pulsational mode of a star from the Q value alone. Too many non-radial
modes are possible.}

We carried out the analogous calculations for BS 1319 and BS 1459 using
Str$\rm \ddot{o}$mgren photometry from Hauck \& Mermilliod (1990).  The
bolometric correction for F5 V stars is -0.04 (Harris 1963).  For BS 1319
we find $Q = 0.965$ d.  Stars pulsating in the fundamental radial mode can
have $Q$ values as high as 0.12 d (Cox 1980).  {\em If} BS 1319 is pulsating,
it must be exhibiting non-radial $g$-modes.

If $f$ = 6.0 d$^{-1}$ is a true frequency of pulsation for BS 1459, its
pulsation constant is $Q = 0.113$ d.  If $f$ = 18.0 d$^{-1}$
is instead a true frequency of pulsation (see Fig 4), then $Q = 0.037$ d, and
BS 1459 might possibly be a $\delta$ Scuti star.  Given that BS 1459 lies
far outside the cool edge of the Cepheid instability strip, this would
indeed be remarkable.  No firm conclusions should be made concerning its
behavior unless and until its variability is confirmed.

\section{Discussion}

We have shown that the stars BS 1354, BS 1385, BS 1408, BS 1430, BS 1432,
and BS 1472 in the Hyades cluster are, for all intents and purposes,
constant in brightness.  BS 1459 might be photometrically variable at the
$\pm$2 mmag level, but it is likely constant as well.  These seven stars,
along with the comparison star we used, BS 1422, can be considered
excellent photometric standards.

Since some $\gamma$ Dor stars are found outside the cool edge of the Cepheid
instability strip in the HR Diagram, it could be that BS 1319 is a
low-amplitude $\gamma$ Dor star.  However, the near coincidence of the
photometric period and the upper limit to the rotational period would favor
the explanation that its variability is due to star spots.

We have provided some evidence that BS 1459 might be a low-amplitude
$\delta$ Scuti star, but we consider this an unlikely possibility.  The
variability of BS 1459 should be confirmed before any strong claims are
made along these lines.

To substantiate the tentative conclusions obtained here concerning the
variability of BS 1319 and BS 1459 would require much more extensive
photometric monitoring of comparably high precision.  In the case of BS
1319 multi-site photometry would be required to avoid aliasing in the power
spectrum.  BS 1459 might be adequately studied from a single site, given its
much shorter implied period.

The most important result of this paper is that we find no early-F stars
in the Hyades that clearly exhibit behavior like $\gamma$ Dor, 9 Aur, HD
224638 and HD 224945.  $\gamma$ Dor itself is presumably a young object,
since it is imbedded in a $\beta$ Pictoris-like disk or envelope (Walker \&
Wolstencroft 1988).  The Pleiades (having age $\approx$ 78 Myr; Lang 1992)
contains 4 $\delta$ Scuti stars listed by Rodriguez et al. (1994) and one
$\gamma$ Dor candidate (Breger 1972; Krisciunas \& Handler 1995).  The cluster
NGC 2516 contains a number of $\delta$ Scuti stars and also a number of
$\gamma$ Dor candidates.  Snowden (1975) claims that NGC 2516 has an age of
137 Myr, or in any case is more evolved than the Pleiades.  The Hyades
(age $>$ 600 Myr) contains 8 $\delta$ Scuti stars, but apparently no
$\gamma$ Dor stars.  Present evidence lets us hypothesize that
the $\gamma$ Dor phenomenon is a characteristic of young F-type stars.
Strong evidence of this will require many new observations and also
theoretical breakthroughs.

\section*{ACKNOWLEDGMENTS}

We are grateful to University of Hawaii, Institute for Astronomy, for
telescope time on the 0.6-m telescope at Mauna Kea.  KK thanks Malcolm
Smith for director's discretionary time at Cerro Tololo.  KK further
thanks the Joint Astronomy Centre for travel funds.  MR's observing
expenses were paid by a fund generously endowed by William Albrecht.  Some of
the Mauna Kea observations were made with the assistance of P. Pobocik.
Some information for this paper was obtained from the {\sc simbad} data
retrieval system, a data base of the Astronomical Data Centre in
Strasbourg, France.  We thank Michel Breger and Gerald Handler for
a copy of the {\sc perdet} program, and for useful discussions. Luciano
Mantegazza kindly provided software for calculating parameters of interest
from Str$\rm \ddot{o}$mgren photometry.

\section*{NOTE ADDED IN PRESS}

Eggen (1995) has independently suggested that the $\gamma$ Doradus
phenomenon is a characteristic of {\em young} F stars.  He claims that
$\gamma$ Dor itself is part of the IC 2391 supercluster, with an age of
50 Myr.  BS 8799, another $\gamma$ Dor candidate, is a member of the Pleiades
supercluster.  He claims that HD 164615 and 9 Aur are young disk stars, owing
to their space velocities.  HD 224638 and HD 224945 have very small proper
motions, characteristic of young disk stars.  All these stars are to be
found in the ``$\rm B\ddot{o}hm$-Vitense decrement'', a gap in the color-
magnitude diagram of cluster stars.  Stars which have {\em not yet} undergone
the abrupt onset of stellar convection, presumably younger stars, are to be
found in this gap.

\bsp

\end{document}